\documentclass[a4paper,11pt,pdflatex]{article}
\usepackage{pos}
\usepackage{graphicx}

\renewcommand{\logo}{\relax}

\title{High Level Reconstruction with Deep Learning using ILD Full Simulation\footnote{This work was carried out in the framework of the ILD Concept Group.}}
\ShortTitle{High Level Reconstruction with Deep Learning using ILD Full Simulation}

\author*[a]{Taikan Suehara}
\author[a]{Risako Tagami}
\author[b]{Lai Gui}
\author[a]{Tatsuki Murata}
\author[c]{Tomohiko Tanabe}
\author[a]{Wataru Ootani}
\author[a]{Masaya Ishino}

\affiliation[a]{ICEPP and Department of Physics, The University of Tokyo,\\
  7-3-1 Hongo, Bunkyo-ku, Tokyo, Japan}
\affiliation[b]{Physics Department, Imperial College London,\\
London SW7 2AZ, United Kingdom}
\affiliation[c]{MI-6, Ltd.,\\
Nihombashi-Kobunacho 8-13, Chuo-ku, Tokyo, Japan}

\emailAdd{suehara@icepp.s.u-tokyo.ac.jp}

\abstract{
Deep learning can give a significant impact on physics performance of electron-positron Higgs factories such as ILC and FCCee.
We are working on two topics on event reconstruction to apply deep learning.
The first is jet flavor tagging, in which we apply particle transformer to ILD full simulation
 to obtain jet flavor, including strange tagging.
The second is particle flow, which clusters calorimeter hits and assigns tracks to them to improve jet energy resolution.
We modified the algorithm developed in context of CMS HGCAL based on GravNet and Object Condensation techniques and add a track-cluster assignment function into the network.
The overview and performance of these algorithms are described.
}

\FullConference{42nd International Conference on High Energy Physics (ICHEP2024)\\
18-24 July 2024\\
Prague, Czech Republic\\}

\begin{document}
\maketitle

\section{Introduction}

Event reconstruction at collider detectors is a essential task to interpret signals at the detectors into physical objects like particles and jets.
Since recent detector systems have high segmentation, the reconstruction algorithms for them need to be more sophisticated to obtain maximal information with intelligent pattern recognition,
which is suitable for modern machine learning (ML).

In this paper, we present two key algorithms of event reconstruction for Higgs factories; one is jet flavor tagging and the other is Particle Flow Algorithm (PFA).
Higgs factories such as ILC, FCCee or CEPC are proposed electron-positron colliders to produce many Higgs bosons to improve measurements and searches related to Higgs bosons.
For most of detectors for Higgs factories, the flavor tagging and particle flow are among essential algorithms to maximize physics performance for Higgs bosons and related targets.

\section{Simulation and Data samples}

In this study we utilized the full detector simulation events for International Large Detector (ILD)\cite{ILD:2020qve}, one of ILC detector concepts.
The ILD employs silicon vertex detector, silicon inner and outer tracker and main Time Projection Chamber for the trackers at barrel region
while silicon-only tracking at forward region. The calorimeters consist of high granular sensor elements, with size of 5 mm to 3 cm.
A solenoid with 3.5 Tesla magnetic field is equipped outside the entire calorimeter, with additional muon layers outside the magnet.

The software stack (iLCSoft) consists of Geant4 based detector simulation with the ILD detector setup (DDSim), digitizers emulating
detector effects and a tracking software as low-level reconstruction. For flavor tagging studies, standard Particle Flow Algorithm (PandoraPFA)\cite{Thomson:2009rp} is used to reconstruct
particles, and Durham jet clustering algorithm is used to reconstruct jets. Particles inside each jet are used to produce input variables.
FCCee Delphes simulation sample of IDEA detector design is also used for comparison of the performance.
We used $e^+e^- \to ZH \to \nu\nu{}q\bar{q}$, at 250 GeV (for ILC) or 240 GeV (for FCCee) center-of-mass (CM) energies for performance studies.
For Particle Flow, the reconstructed tracks and digitized calorimeter hits are used as input variables.
We utilize 10 $\tau^-$ particles with random directions ovarlayed on single event as well as jet events by $e^+e^- \to q\bar{q}$ at 91 GeV CM energy.

\section{Flavor tagging with Particle Transformer (ParT)}


Jet flavor tagging is an algorithm to classify jets with jet properties.
In the previous algorithms used for Higgs factory studies, LCFIPlus\cite{LCFIPlusPaper}, the secondary vertices are reconstructed from off-axis tracks, and then jets are classified using
the vertex and additional track features. In contrast, there are deep-learning based algorithms 
directly using all track information into the input of the network. Such algorithms are applied to LHC jet analyses and already give much better results than previous ones.
Among the algorithms, ParticleNet which is based on point clouds and graph neural network is applied to FCCee fast simulation
and nearly 10 times better rejection ratio than LCFIPlus for $b$-tagging has been reported\cite{FCCPN}.
More recently, Particle Transformer (ParT), based on Transformer, reported exceedings performance with LHC dataset.
Transformer is the network used for many language-processing network, and has particularly competitive performance to large training samples.
We apply the ParT to ILD full simulation dataset as well as FCCee fast simulation dataset for comparison.

\begin{figure}[htbp]
\centering
\includegraphics[width=0.8\textwidth]{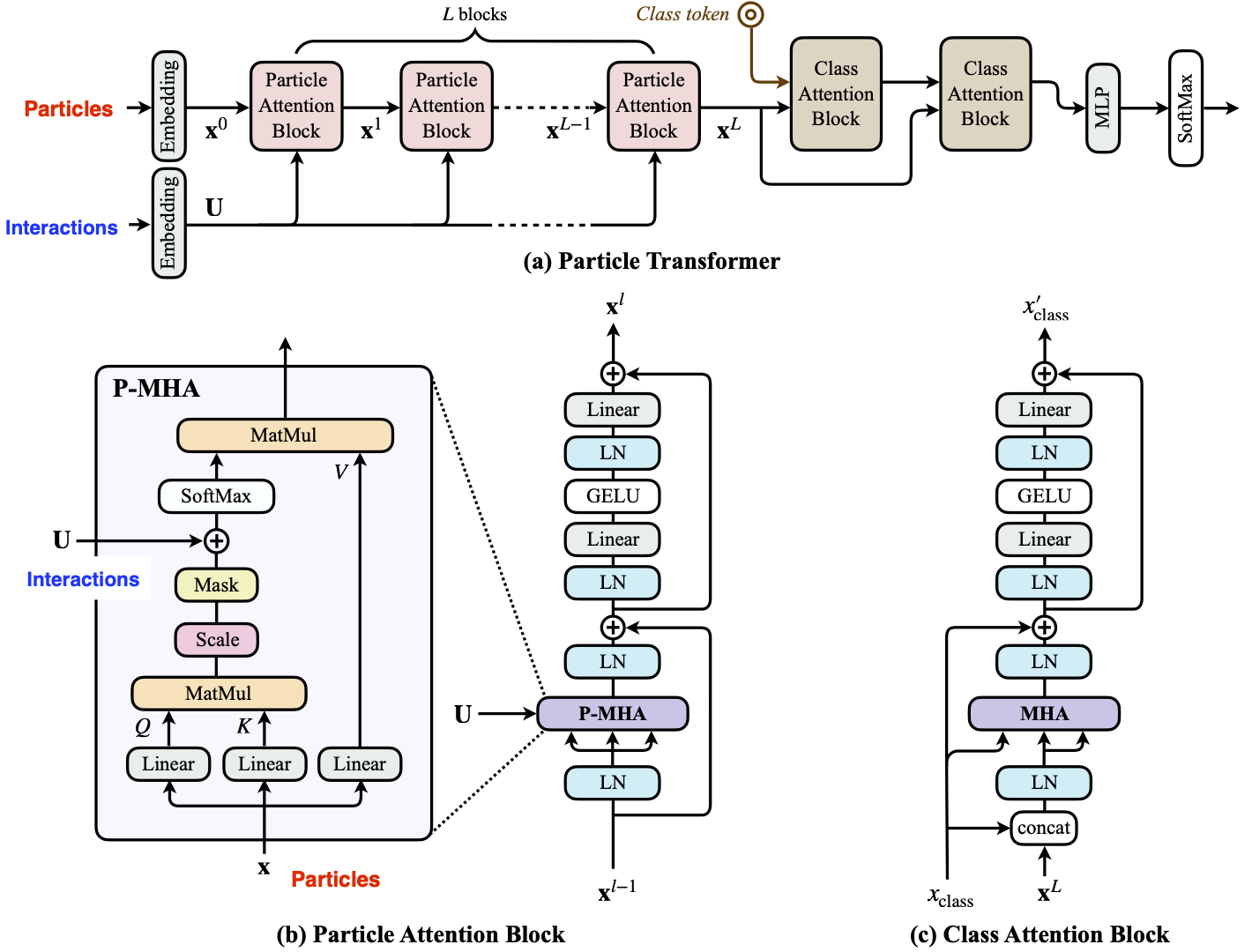}
\caption{Schematic diagram of Particle Transformer (ParT) \cite{Qu:2022mxj}.}
\label{fig:part}
\end{figure}

Figure \ref{fig:part} shows the schematic diagram of ParT. It consists of embedding layers, self-attention layers and fully-connected layers as similar to plain Transformer,
but the attention weights are calculated not only from self-attention layers but output from ``interaction" part is added as a bias to the weights.
Charged and neutral particles are treated as distinct input elements with separate embedding layers.
For charged particles, features are impact parameters, track errors, particle ID variables and kimematic variables while
neutral particles only have kinematic and particle ID variables. The interaction is calculated from 4-momentum of the two particles to calculate attention weights.
Features of the interaction is kinematic variables of two particles such as invariant mass.
Details of input variables are summarized in \cite{lcws2024-tagami}.

\begin{table}[htb]
 \centering
\small
  \begin{tabular}{l r r r r} \hline
    configuration  & \multicolumn{2}{c}{$b$-tag 80\% eff.} & \multicolumn{2}{c}{$c$-tag 50\% eff.} \\
    background     & $c$-bkg. & $uds$-bkg. & $b$-bkg. & $uds$-bkg. \\ \hline
    ILD full sim 1M (LCFIPlus)        & 6.3\%  & 0.79\% & 7.4\%  & 1.2\% \\
    ILD full sim 1M (ParT)            & 0.48\% & 0.14\% & 0.86\% & 0.34\% \\
    FCCee Delphes 1M (ParT, reduced)  & 0.47\% & 0.12\% & 0.64\% & 0.10\% \\
    FCCee Delphes 1M (ParT, full)     & 0.21\% & 0.054\% & 0.36\% & 0.059\% \\
    FCCee Delphes 4M (ParT, full)     & 0.045\% & 0.025\% & 0.20\% & 0.033\% \\
    FCCee Delphes 6M (ParT, full)     & 0.014\% & 0.010\% & 0.13\% & 0.022\% \\
    FCCee Delphes 8M (ParT, full)     & 0.007\% & 0.006\% & 0.076\% & 0.021\% \\ \hline
  \end{tabular}
  \caption{Background acceptance on 3-category flavor tagging, compared between ILD and FCCee and sample size with 250/240 GeV $\nu\nu{}qq$ sample. FCCee with reduced results is trained with partial input parameter set which is nearly compatible with ILD. Full dataset includes more input variables.}
  \label{tbl:3cat}
\end{table}

Table \ref{tbl:3cat} shows the performance of the 3-category ($b$, $c$, and others) flavor tagging comparing LCFIPlus, ParT with ILD and ParT with FCCee with various configurations.
It shows factor 5-10 improvements on rejection of jets by ParT from LCFIPlus, and ILD and FCCee results are not significantly different if we use compatible variables.
The FCCee results also show that the performance is sensitive to the sample size of the training, which needs to be confirmed with full simulation.

For 6-category ($b$,$c$,$s$,$u$,$d$, and $g$) identification, it is essential to include and optimize the particle identification (PID) of the tracks since existence of high-momentum Kaons
is one of the critical measure to distinguish strange jets from others.
We utilize Comprehensive PID (CPID)\cite{cpid} for ILD full simulation, which is based on BDT with 12 momentum bins, using dE/dx by TPC and time-of-arrival (ToF) obtained
with 10 track-like calorimeter hits with 100 psec timing resolution assumed for individual hits.
Probabilities of proton, kaon, pion, muon and electron are used as input features of ParT.
We have also provided results with truth PID, which cheats truth PID information instead of using probabilities.
For FCCee fast simulation, dN/dx, which is cluster counting per unit length and track mass calculated from ToF are directly used as input features of ParT
instead of using probabilities.
Table \ref{tbl:stag} shows the performance of the strange tagging.
We can expect some statistical power of the separation of strange tag with ILD full simulation, while the performance is worse than FCCee result.
The difference between realistic and truth PID gives part of the explanation of the difference, but since FCCee result is still better than ILD with truth PID,
there should be unknown reason, which needs to be investigated. 


\begin{table}[htb]
 \centering
\small
  \begin{tabular}{l r r} \hline
    configuration  & \multicolumn{2}{c}{$s$-tag 80\% eff.}  \\
    background     & $g$-bkg. & $d$-bkg. \\ \hline
    ILD full with CPID          & 25.7\% & 42.7\%  \\
    ILD full with truth PID     & 23.2\% & 38.0\%  \\
    FCCee Delphes 1M            & 20.3\% & 29.6\%  \\ \hline
  \end{tabular}
  \caption{Background acceptance on 6-category strange tagging, compared between ILD and FCCee  with 250/240 GeV $\nu\nu{}qq$ sample.}
  \label{tbl:stag}
\end{table}

\section{Particle flow with DNN}

Particle Flow Algorithm (PFA) is an algorithm to reconstruct particles from tracks and calorimeter clusters for high-granular calorimeters,
consisting of hit clustering and track-cluster association. Accuracy of the track-cluster assignment is critical for the jet energy resolution with PFA.

\begin{figure}[htbp]
 \centering
 \begin{minipage}[t]{0.6\hsize}
  \centering
  \includegraphics[width=1.0\textwidth, clip]{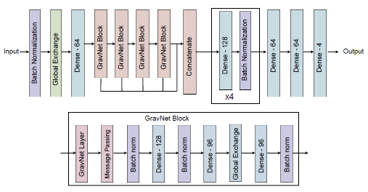}
  \caption{Structure of our PFA network\cite{Qasim:2021hex}.}
  \label{fig:gnn-structure}
 \end{minipage}
 \begin{minipage}[t]{0.35\hsize}
  \centering
  \includegraphics[width=0.9\textwidth, clip]{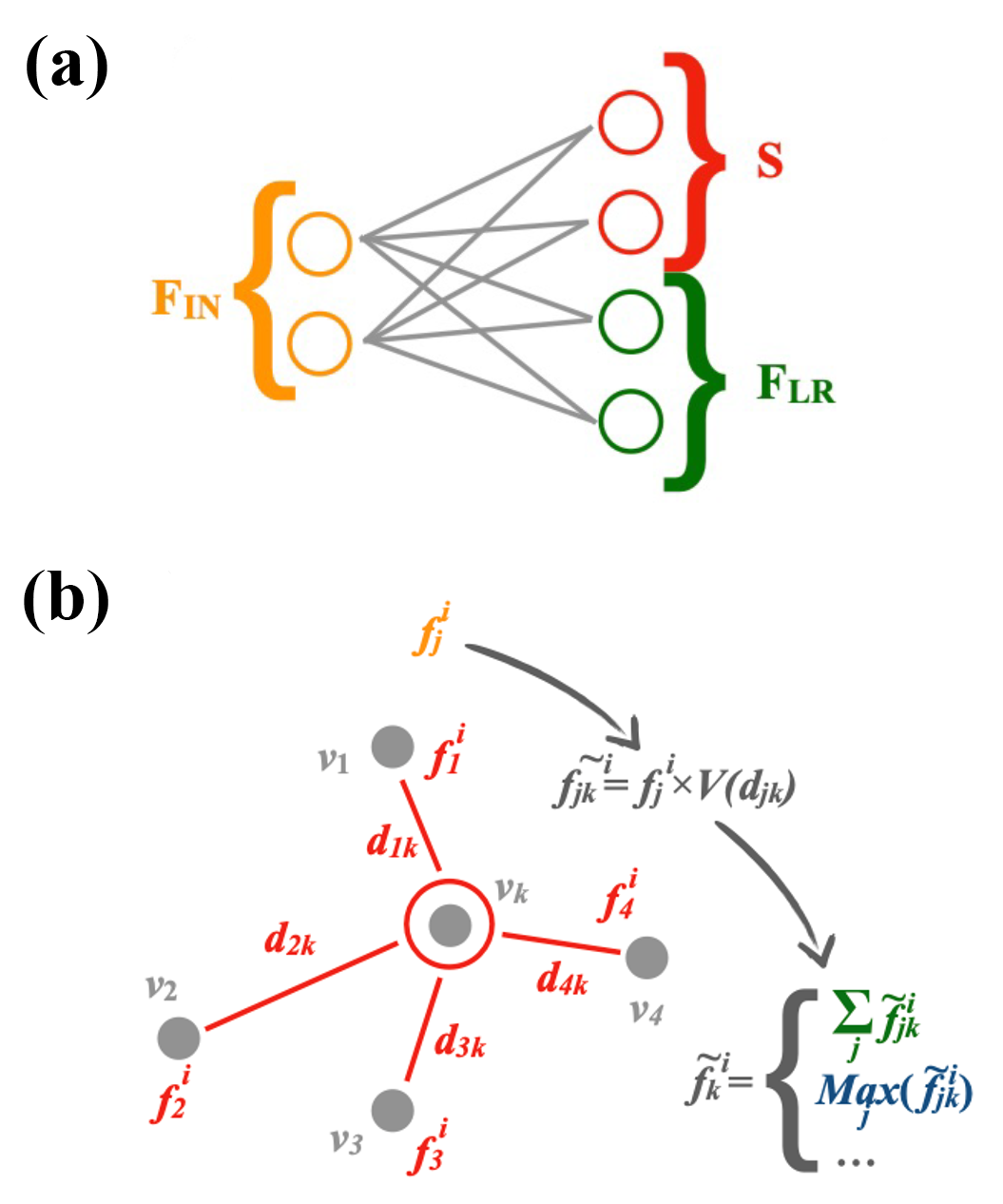}
  \caption{Schematic of GravNet. }
  \label{fig:gravnet}
 \end{minipage}
\end{figure}

In this study we utilize a Graph Neural Network (GNN)-based method, GravNet, for clustering and track association.
Figure \ref{fig:gnn-structure} shows schematic view of our network\cite{Qasim:2021hex}. Our input is calorimeter hits and track information and output is a virtual coordinate
 and ``condensation parameter" $\beta$ which is obtained with ``object condensation" loss function.
We utilize position and energy deposit of the hits as input features. For tracks, position where the track enter the calorimeter is used
as the position and energy deposit is set to zero. We also have a flag to separate normal hits and tracks.

GravNet is a distance-based GNN algorithm. The initial features are combined into virtual coordinates (S) and other features (F$_\mathrm{LR}$)
using a simple MLP as shown in Fig.~\ref{fig:gravnet}(a), then do convolution with neighbor nodes using the virtual coordinate as shown in Fig.~\ref{fig:gravnet}(b).
The convoluted nodes are used for the second round, and the output of each stage as well as initial feature are concatenated to obtain the output of the network.
The object condensation is a loss function described as
$L = L_\beta + L_V$
where $L_\beta$ is a term to pull up $\beta$ of one hit while putting down $\beta$ of other hits,
and $L_V$ consists of attractive potential to a condensation point of the same truth cluster and repulsive potential to the
condensation point of the different truth cluster. The condensation point is defined as the hit with highest $\beta$ if
no tracks are associated. If there exists a track inside the truth cluster, the track is treated as the condensation point.

We apply a clustering algorithm from the output of the network to obtain particles.
In our method, all hits and tracks having more than $t_\beta$ are treated as condensation points
and hits within distance of $t_d$ at the virtual output coordinate from each condensation point are clustered into the condensation point.
$t_\beta$ and $t_d$ are tunable parameters which we scanned to obtain best results.

For the evaluation of the performance, we defined sum of energy deposit of truth cluster as $e_t$, sum of energy deposit of
predicted cluster as $e_p$ and sum of energy deposit of hits in a reconstructed cluster which are coming from the same truth cluster
as $e_m$. Efficiency is defined as sum of $e_m/e_t$ for all truth clusters and purity is defined as $e_m/e_p$.
The efficiency and purity are avaraged over all clusters in the event sample to evaluate the performance of the clustering.
The parameters $t_\beta$, $t_d$ and dimension of the output coordinate are tuned by looking efficiency-purity plane.

\begin{table}[htb]
 \centering
\small
  \begin{tabular}{l r r r r r r} \hline
    algo/events  & Electron eff. & Pion eff. & Photon eff. & Electron pur. & Pion pur. & Photon pur.\\ \hline
    GravNet/Taus    & 99.1\% & 96.5\% & 99.0\% & 91.8\% & 98.9\% & 97.1\% \\
    PandoraPFA/Taus & 99.3\% & 94.0\% & 99.1\% & 91.8\% & 94.6\% & 97.2\% \\
    GravNet/Jets    & 94.5\% & 93.1\% & 95.2\% & 77.4\% & 93.2\% & 92.4\% \\
    PandoraPFA/Jets & 80.2\% & 90.4\% & 79.0\% & 75.0\% & 90.6\% & 77.2\% \\ \hline
  \end{tabular}
  \caption{Performance on efficiency and purity with our GravNet-based algorithm compared with PandoraPFA.}
  \label{tbl:pfa}
\end{table}

Table \ref{tbl:pfa} shows the performance of the efficiency and purity of our method as well as PandoraPFA.
We clearly see improvements on both efficiency and purity of pions by our new method for both tau and jet samples.
Performance on jet energy resolution needs energy regression of the clusters, which is now under investigation.


\section{Summary and plans}

We studied two critical reconstruction algorithms, jet flavor tagging and Particle Flow Algorithm with modern deep learning technologies\footnote{This work is supported by JSPS KAKENHI Grant Numbers JP22H05113, JP23H04513.}.
The flavor tagging with Particle Transformer shows almost one-order-of-magnitude improvement of rejection of the different jet flavors
with fixed efficiency on $b$ and $c$ tagging, and non-negligible separation of strange jets. 
For the particle flow, our DNN-based method gives superior performance on efficiency and purity of the clustering compared to
old method, PandoraPFA, but still working on energy regression to achieve better jet energy resolution.

We are also developing completely different method of PFA which is based on a Transformer-based algorithm with 
encoder-decoder framework which is widely used for natural language processing\cite{lcws2024-paul}.
 This method is expected to have
more flexibility on parameters, and also having a hope to apply for related problems like jet clustering with the same network.
Based on these developments, we would also aim for full DNN-based reconstruction chain in mid-term future.

\bibliographystyle{JHEP}
\bibliography{ichep2024-suehara}

\end{document}